\documentstyle[preprint,prb,aps]{revtex}

\begin{document}
\draft
\title{Understanding light quanta:\\
The Photon}
\author{Alberto C. de la Torre }
\address{Departamento de F\'{\i}sica,
 Universidad Nacional de Mar del Plata\\
 Funes 3350, 7600 Mar del Plata, Argentina\\
dltorre@mdp.edu.ar}
\maketitle
 \begin{center}
quant-ph/0410179 \end{center}
\begin{abstract}
An antisymmetric tensor, the photon tensor, is defined for the
description of the photon as a massless relativistic particle.
The classical photon can then be visualized as an essentially two
dimensional rotating object. The quantum mechanical description
of a single photon is presented and the wave-particle duality for
the photon is analysed. The intended relation between
Schr{\"o}dinger's equation for the photon with Maxwell's equations is
discussed. This work is devoted to the attempt to understand the
quantum of electromagnetic radiation, based on the assumption
that the photons are the primary ontology and that the
electromagnetic fields are macroscopic emergent properties of an
ensemble of photons.
\end{abstract}
\section{INTRODUCTION}
On March 2005 we celebrated the one hundredth anniversary of the
photon. At the beginning, the particle character of the photon
was not recognized and it was considered to be a parcel of
electromagnetic energy\cite {ei}. Only after his twentieth
anniversary and after the discovery of the Compton
effect\cite{compton}, the photon was accepted as a full fledged
particle and he was given a name\cite{lewis}. Even though the
concept of the photon is old, there are still many confusing
matters that need clarification and there are opposing
interpretations concerning the reality of the photon; indeed,
near the end of his live, Einstein complained that after fifty
years of conscious meditation he had not come any closer to the
answer to the question \emph{What are light quanta?}\cite{ei2}.

In a previous work\cite{foton1}, the commutation relations of the
operators corresponding to the electromagnetic fields were
calculated from first principles, without reference to quantum
field theory, and their highly singular character was discussed.
The conclusion reached was that we should not consider the
electromagnetic fields to be a primary ontology but that they are
a macroscopic collective effect of a large, or undetermined,
number of particles, the photons. One should however mention that
the choice adopted, namely, that \emph{the photons are the
primary ontology that must be treated quantum mechanically and
that the electromagnetic fields are macroscopic manifestations of
an ensemble of them}, is not a unique choice. There are indeed
authors that consider the electromagnetic fields as the primary
ontology and that the photons do not have an objective existence
but are rather mathematical entities, ``particle-like
excitations''\cite{bal}, corresponding to the normal mode
decomposition of the fields. This alternative option can be
adopted because, although the objective existence of the photons
provide the most natural explanation of the photoelectric effect
and of Compton scattering, these are not compelling evidence for
their existence because these effects can also be explained by a
semiclassical theory combining unquantized electromagnetic fields
with quantum theory of
matter\cite{semiclas1,semiclas2,semiclas3}. These semiclassical
theories are however not capable of explaining some phenomena
like the spontaneous emission of light by atoms or the
correlations between distant photons in Clauser-Aspect test of
Bell's inequalities.

What is the meaning of trying to understand the photon? We know
that the photon carries energy-momentum and has spin.  Are these
two things related? We can argue that they can not be related
because spin has always the same value for all photons -one unit
of $\hbar$- whereas the energy can vary continuously from zero to
infinity. On the other side, spin is apparently the only property
possessed by the photon and we don't know what in the reality of
the photon is the thing that carries energy. We expect that the
energy of the photon  resides in some internal property of it and
is not purely kinematic because all photons move identically,
with the same speed, but they can have different energies. What
differentiates two photons of different energy? We will see that,
as a consequence of \emph{special relativity}, not of quantum
mechanics, the energy of a photon is given by some frequency.
What is the thing that is changing in time with such a frequency?
Is something within the photon rotating with that frequency?
Wouldn't it be natural to associate this rotation with the spin
of the photon? All these questions and many other, indicate that
we are still lacking a deep understanding of the quanta of
electromagnetic radiation and although some progress has been
achieved, Einstein's question \emph{What are light quanta?} has
not received a complete answer.

In this work we will present the quantum mechanical treatment of
the photon  in a ``first quantization'' style, similar to a
previous presentation\cite{birul}, but with the difference that
here we do not rely on Maxwell's equations as a staring point but
instead we define the photon as a particle described with a
``photon tensor''. We find this way more convenient because there
is some misunderstanding relating the quantum mechanical state of
a photon, solution of the photon's Schr{\"o}dinger equation,  with
the electromagnetic fields, solutions of Maxwell's equations. We
will show, indeed, that the intended derivation of Maxwell's
equation from Schr{\"o}dinger equation is not justified. We will see
that, in this interpretation, the photon is not the particle-like
duality partner of the wave associated to the classical
electromagnetic field. The particle-wave duality in the
electromagnetic radiation corresponds to two different quantum
states of the photon. On one side the space localized states
$\varphi_{{\mathbf r}}$, eigenvectors of the position operator
${\mathbf R}$, and on the other side, states with sharp momentum
and energy $\phi_{{\mathbf p}}$. In both cases we deal with
single photon states and not with the electromagnetic field
resulting from the combined effect of an undetermined number of
photons. The confusion arises, probably, because the classical
electromagnetic plane wave is closely related to a Bose-Einstein
condensate of a large and indefinite number of photons in one
single state $\phi_{{\mathbf p}}$.

Greek indices $\mu,\nu,\sigma,\cdots$ assume values $0,1,2,3$ and
the $0$ is associated with the time coordinate; latin indices
$i,j,k,\cdots$ take values in $1,2,3$; contravariant vectors will
be occasionally decomposed in their time and space parts
$k^{\mu}=( k^{0},{\mathbf k})$; the metric tensor is
$g^{\mu\nu}=g_{\mu\nu}=$Diag$(1,-1,-1,-1)$;
 $\varepsilon_{\mu\nu\sigma\rho}$ and $\varepsilon_{jkl}$ are the total antisymmetric
tensor, $\partial_{k}= \frac{\partial}{\partial x_{k}}$,
$\partial_{t}= \frac{\partial}{\partial t}$ and we adopt
Einstein's convention, stating that repeated indices indicate a
summation.
\section{THE CLASSICAL PHOTON}
Let us postulate the existence of a massless physical system,
called \emph{photon}, transporting energy $E$, momentum
$\mathbf{P}$ and intrinsic angular momentum (spin) $\mathbf{S}$ .
The relativistic kinematics of a massless particle implies that
its speed must be $c$ (Einstein's constant) and that the energy,
momentum and spin are related by
\begin{equation}\label{ep}
 E=c|{\mathbf P}|
\end{equation}
and
\begin{equation}\label{constr}
 {\mathbf S}\times {\mathbf P} =0\ .
\end{equation}
The constraint imposed by the second relation, that for reasons
to be clarified later can be called \emph{transversality}
constraint, follows from the fact that the intrinsic angular
momentum of an extended object moving at the speed $c$ can not
have any component perpendicular to the direction of propagation.
This can be easily understood if we consider a sphere moving with
speed $v$ \emph{less} than $c$ in a direction ${\mathbf k}$ and
rotating around an axis along ${\mathbf s}$, perpendicular to the
direction of propagation. Two points on the periphery ``above''
and ``below'' the plane of ${\mathbf k}$ and ${\mathbf s}$ will
have different velocities  resulting from the relativistic
combination of the velocity $v$ of the center with the tangential
velocity $\pm v_{T}$ due to the rotation. Now, if the center
moves with speed $c$, the two points on the periphery will also
have the speed $c$ and no rotation is possible. Therefore an
extended massless object moving with speed $c$ can only rotate
along an axis collinear with the direction of propagation as
required by Eq.\ref{constr}. This heuristic argument would not be
valid for a point like particle without spatial extension;
however the quantum nature of particles suggest that no particle
is exactly point like: massive particles can not be localized
beyond their Compton wave length and a photon moving in a well
defined direction must be extended in a direction perpendicular
to the direction of movement as imposed by Heisenberg's
uncertainty principle.

The massless character of the photon suggests that its energy
must be proportional to some frequency; that is, \emph{something
in the photon must be changing periodically in time}. The
energy-frequency relation is therefore a consequence of special
relativity. Although this was known for long\cite{rind}, it has
not received much attention. In order to see this, let us
consider a photon propagating in the $x^{1}$ direction with four
momentum (with $c=1$) $p^{\mu}=(E,E,0,0)$. We can now perform a
Lorentz transformation with speed $\beta$ (Lorentz factor
$\gamma=1/\sqrt{1-\beta^{2}}$) in the direction of the $x^{1}$
coordinate axis. Doing this we see that the energy $E$ of the
photon decreases to $E'$ in the ``primed'' reference frame given
by
\begin{equation}\label{frecu}
  E'=E\:\gamma(1-\beta)=E\:\sqrt{\frac{1-\beta}{1+\beta}} \ .
\end{equation}
However this is precisely the transformation property of a
frequency (Doppler shift) and we can therefore expect that the
energy of a photon depends on some frequency. What makes this
most remarkable, is that we have found that the energy of the
photon depends on its frequency using only classical relativistic
(that is, non-quantum) arguments in contrast with usual saga that
presents the energy of a photon $h\nu$ as an essential quantum
mechanical fact. The energy-frequency relation was discovered by
Einstein in his seminal photoelectric effect paper\cite{ei} in
1905, and is usually publicized as one of the first quantum
postulates. However, after the consolidation of the photon
concept as a particle this relation becomes a consequence of the
massless character of any relativistic particle and, by the way,
the same should hold for gluons and gravitons.

In order to build our model of a photon we must decide
\emph{what} is the thing or quality that carries energy and
momentum. For an electron, we would think on a electrically
charged massive particle carrying one half unit of angular
momentum and possibly other qualities such as the leptonic charge
and so on. We must then postulate the existence of some
``elements of physical reality'', as Einstein could have called
it, in terms of which, the energy and spin of the photon should
be expressed. Einstein's relativity requires that this elements
of physical reality should be formalized by a mathematical entity
having well defined properties under a Lorentz transformation
relating different observers, that is, it should be a scalar,
vector or tensor of appropriate rank. Clearly, a scalar, being an
invariant, can not represent the properties of a photon and
neither can we describe the photon with a vector (different from
$p^{\mu}$). We can then consider an antisymmetric, or symmetric
or general second rank tensor with 6, 10, 16 components
respectively. We have no profound reason to decide, except that
if we choose the simplest case, an antisymmetric tensor
$f^{\mu\nu}$ in order to describe such an element of physical
reality, then we can accommodate all known physical properties of
the photons. Let us choose the name \emph{``Photon Tensor''} in
order to denote this antisymmetric tensor that, therefore,
contains six independent real numbers. It is very convenient to
describe these six real numbers in terms of two three-vectors
${\mathbf e}$ and ${\mathbf b}$. However we should warn that, in
rigour, the quantities ${\mathbf e}$ and ${\mathbf b}$ \emph{are
not} the space components of a four-vector because they do not
transform as such in a general Lorentz transformation. They
should be considered as a convenient notation for the six
nonvanishing components of the photon tensor $f^{\mu\nu}$,
according to the assignment given by
\begin{equation}\label{fmunu}
   f^{\mu\nu}= \left(%
\begin{array}{cccc}
  0 & e_{1} & e_{2} & e_{3} \\
  -e_{1} & 0 & b_{3} & -b_{2} \\
  -e_{2} & -b_{3} & 0 & b_{1} \\
  -e_{3} & b_{2} & -b_{1} & 0 \\
\end{array}%
\right)\ .
\end{equation}
The notation is anyway justified because ${\mathbf e}$ and
${\mathbf b}$ transform as three-vectors under the
\emph{subgroup} of Lorentz transformations corresponding to space
rotations and translations.

Associated to any second rank antisymmetric tensor there is
another second rank antisymmetric tensor, the dual tensor,
defined by
\begin{equation}\label{fmunustsar}
   f^{\star\:\mu\nu}=\frac{1}{2}\varepsilon^{\mu\nu\sigma\rho}f_{\sigma\rho}=
\left(%
\begin{array}{cccc}
  0 & -b_{1} & -b_{2} & -b_{3} \\
  b_{1} & 0 & e_{3} & -e_{2} \\
  b_{2} & -e_{3} & 0 & e_{1} \\
  b_{3} & e_{2} & -e_{1} & 0 \\
\end{array}%
\right)\ .
\end{equation}
Notice that the dual photon tensor is obtained from the photon
tensor performing a \emph{dual transformation} defined by
${\mathbf e}\rightarrow{-\mathbf b}$ and ${\mathbf
b}\rightarrow{\mathbf e}$. The description of the photon by means
of the photon tensor is equivalent to its  description with the
dual photon tensor and we postulate that the photon is invariant
under the dual transformation.

 Before we start with the physical analysis of the photon
tensor it is important to clarify that, in spite of the formal
similarity, the photon tensor $f^{\mu\nu}$ \emph{is not a field}
like the electromagnetic field tensor $F^{\mu\nu}(t,{\mathbf
r})$. The components of $f^{\mu\nu}$ are not functions of
space-time and therefore derivatives like
$\partial_{\mu}f^{\mu\nu}$ are meaningless. There are, indeed,
many formal differences between the photon tensor and the
electromagnetic tensor. It is therefore wrong to think about
$f^{\mu\nu}$ as being ``the electromagnetic field of one photon''
because, in this interpretation, the electromagnetic field is an
emergent property of an \emph{ensemble} of photons; the idea of
the electromagnetic field of \emph{one} photon is as meaningless
as the idea of the density or temperature of \emph{one} molecule.

The three scalars associated with the photon tensor are
\begin{eqnarray}
 f^{\mu}_{\mu}&=& 0\ , \\
  f^{\mu\nu}f_{\mu\nu}&=& 2\ (e^{2}-b^{2})\ ,\\
  f^{\mu\nu}f^{\star}_{\mu\nu}&=&
-4\ {\mathbf e}\cdot{\mathbf b}\ ,
\end{eqnarray}
where $e=|{\mathbf e}|$ and $b=|{\mathbf b}|$. The first scalar
vanishes trivially and we will later see that the other two
scalar also vanish.

Since every proper orthochronous Lorentz transformation is
equivalent to a standard Lorentz transformation preceded and
followed by an appropriate space rotation and translation, we
only need to consider the transformation of the photon tensor
under a boost in the $x^{1}$ coordinate axis. Doing this we
obtain the photon tensor in the ``primed'' reference frame that
is moving with speed $\beta$ (Lorentz factor
$\gamma=1/\sqrt{1-\beta^{2}}$) along the $x^{1}$ axis with
respect to the unprimed frame:
\begin{equation}\label{fprimmunu}
  f'^{\:\mu\nu}= \left(%
\begin{array}{cccc}
  0 & e_{1} & \gamma (e_{2}-\beta b_{3}) & \gamma (e_{3}+\beta b_{2}) \\
  -e_{1} & 0 & \gamma(b_{3}-\beta e_{2}) & -\gamma(b_{2}+\beta e_{3}) \\
 -\gamma (e_{2}-\beta b_{3}) & -\gamma(b_{3}-\beta e_{2}) & 0 & b_{1} \\
  -\gamma (e_{3}+\beta b_{2}) & \gamma(b_{2}+\beta e_{3}) & -b_{1} & 0 \\
\end{array}%
\right)\ ,
\end{equation}
where we can read the components of ${\mathbf e'}$ and ${\mathbf
b'}$ in the primed reference frame. We can notice here that,
indeed, ${\mathbf e}$ and ${\mathbf b}$ do not transform as the
space components of a four-vector.

The photon tensor is not invariant under space rotations. This
means that the photon has some intrinsic orientation or
directionality, manifest in the tree-vectors ${\mathbf e}$ and
${\mathbf b}$. On the other side, the photon propagates in space
along a direction defined by a unit vector ${\mathbf k}$. These
two directional properties of the photon, the intrinsic
orientation of $f^{\mu\nu}$ and the direction of propagation
${\mathbf k}$, are not independent but are coupled as consequence
of special relativity. Because of the Lorentz contraction for an
object moving with speed $c$, all lengths in the direction of
propagation must collapse to zero; therefore we can think about
the photon as being an \emph{essentially} two dimensional object
with all its elements of physical reality in a plane orthogonal
to the direction of propagation, that is, we must have ${\mathbf
k}\cdot{\mathbf e}=0$ and ${\mathbf k}\cdot{\mathbf b}=0$.
However, these two conditions must be expressed in a way valid in
all reference frames, that is, in a covariant way under Lorentz
transformations. For this we notice that, due to the massless
character of the photon, the unit vector ${\mathbf k}$ must be
the space part of a propagation null four-vector, tangent to the
photon world line, given by $k^{\mu}=( 1,{\mathbf k})$. The
coupling of the intrinsic and external orientations of the photon
becomes then
\begin{equation}\label{fk}
  k_{\mu} f^{\mu\nu} =({\mathbf k}\cdot{\mathbf e}\ ,\ {\mathbf e}+
{\mathbf k}\times{\mathbf b}) = 0 \ ,
\end{equation}
and the dual transformation of this relation that must also be
true,
\begin{equation}\label{fstark}
   k_{\mu}f^{\star\:\mu\nu} =(-{\mathbf k}\cdot{\mathbf b}\ ,\ -{\mathbf
b}+ {\mathbf k}\times{\mathbf e}) = 0 \ .
\end{equation}
Therefore we have
\begin{eqnarray}
 {\mathbf k}\cdot{\mathbf e}=0\ &,&\ {\mathbf k}\times{\mathbf e}={\mathbf b} \\
{\mathbf k}\cdot{\mathbf b}=0\ &,&\ {\mathbf k}\times{\mathbf
b}=-{\mathbf e} \ .
\end{eqnarray}
 From these (redundant) equations it follows that in every Lorentz
frame, ${\mathbf e}$ and ${\mathbf b}$ lay in a plane orthogonal
to the direction of propagation ${\mathbf k}$, that they are
orthogonal and with equal modulus, and $({\mathbf e},{\mathbf
b},{\mathbf k})$ build a right handed  set  of orthogonal
vectors. Therefore we have
\begin{equation}\label{eb0}
    {\mathbf e}\cdot{\mathbf b}= 0\ ,\ e=b \ .
\end{equation}
This result, valid in all reference frames, imply that the two
scalar in Eqs.(7,8) vanish, as mentioned before.  Notice that
there is an arbitrary orientation of the orthogonal pair
$({\mathbf e},{\mathbf b})$ in their plane; therefore there is a
reference frame, with the $x^{1}$ axis along the direction of
propagation, where the photon tensor of Eq.(\ref{fmunu}) assumes
the simplest form with all components vanishing except for
$e_{2}$ and $b_{3}$ that are equal.

In order to relate the energy of the photon with the components
of $f^{\mu\nu}$, let us consider that the energy is invariant
under space rotations; therefore it must depend on rotationally
invariant quantities like $e=|{\mathbf e}|$ or $b=|{\mathbf b}|$.
Furthermore we expect the energy to be invariant under the
duality transformation. However, since $e=b$, we only have to
consider the dependence of the energy from $e$ in a way that
should have the appropriate transformation property given by
Eq.(\ref{frecu}). We must therefore calculate the transformation
property of $e$ for a photon moving along $x^{1}$ under a
standard Lorentz transformation. In this case we have
$e_{1}=b_{1}=0$ and considering the components of the transformed
photon tensor given in Eq.(\ref{fprimmunu}) we have
\begin{equation}\label{etransf}
  e'=\sqrt{\gamma^{2} (e_{2}-\beta b_{3})^{2} + \gamma^{2} (e_{3}+\beta
b_{2})^{2}} =\gamma \sqrt{e^{2}+\beta^{2} b^{2} -
2\beta(e_{2}b_{3}-e_{3}b_{2})}\ .
\end{equation}
Notice that the term in parenthesis is the component of ${\mathbf
e}\times{\mathbf b}$ along the $x^{1}$ axis; therefore using
Eq.(\ref{eb0}) we get
\begin{equation}\label{etransf1}
  e'=e\:\gamma \sqrt{1+\beta^{2}-
2\beta}=e\:\gamma (1-\beta) \ .
\end{equation}
Comparing this with Eq.(\ref{frecu}) we see that the energy of
the photon $E$, and the modulus of the photon tensor $e=b$, have
identical transformation property, that is, they both transform
as a frequency. Then, the energy can be given by an homogeneous
function of $e$ of first degree, that is, $E$ must be
proportional to $e$, and we introduce the symbol $\omega=e=b$ in
order to denote the frequency. The proportionality constant
between the energy and the frequency must have units of action or
of angular momentum and therefore we set it equal to Planck
constant. We have then
\begin{equation}\label{Ehomega}
    E=\hbar\omega\ .
\end{equation}
Both vectors ${\mathbf e}$ and ${\mathbf b}$ have equal modulus,
$\omega$, and we can express them in terms of a set of orthogonal
unit vectors ${\mathbf \hat{e}}$ and ${\mathbf \hat{b}}$ as
${\mathbf e}=\omega{\mathbf \hat{e}}$ and ${\mathbf
b}=\omega{\mathbf \hat{b}}$. Also the photon tensor can have the
frequency $\omega$ factored out and be given in terms of unit
vectors. We can now fix the orientation of the unit vectors
${\mathbf \hat{e}}$ and ${\mathbf \hat{b}}$ in their plane by
requiring that they rotate with the frequency $\omega$. For this
rotation we have two choices corresponding to a clockwise or
counterclockwise rotation. Considering the propagation of the
photon along the direction ${\mathbf k}$, we see that in these
two choices the tip of the unit vector ${\mathbf \hat{e}}$ will
make a right handed or a left handed helix. These correspond to a
positive or a negative helicity photon. The rotation and
propagation of the photon allows us to introduce a useful length
scale for the photon corresponding to the step of the helix given
by
\begin{equation}\label{lambda}
  \lambda= \frac{2\pi c}{\omega}\ .
\end{equation}
Notice that we do not call this length scale ``wave length''
because, so far, for the classical -that is, non quantum- photon
we don't have any ``wave''. It is natural to associate this
rotation of the vectors ${\mathbf \hat{e}}$ and ${\mathbf
\hat{b}}$ with the spin of the photon; however we must remind
that spin can only take one value, $\pm\hbar$, whereas the
frequency (the energy) can take any positive value. This is no
real problem because we can think of a rotating energy
distribution whose associated angular momentum is constant. In
the Appendix we show several examples of these mechanical models.
However we should not take too seriously these mechanical models
of the photon. All that we want to show is that there is no
contradiction between a constant (i.e. for all frequencies) value
of angular momentum of a rotating system and an energy linearly
dependent on the frequency of rotation, therefore we are allowed
to think that spin and energy are due to the rotation of the
element of physical reality of the photon in the plane
perpendicular to the direction of propagation.

The behaviour of the photon tensor under the proper and
orthochronous Lorentz transformations proved to be very useful in
order to determine its classical features. For completeness we
will give now all the transformation properties of the photon
tensor. One of the reasons for characterizing the photon tensor
$f^{\mu\nu}$ in terms of the two three-vectors ${\mathbf e}$ and
${\mathbf b}$ is that these have simpler properties under space
and time inversion and also under charge conjugation and duality
transformation. From the definition of the photon tensor, it
follows that under a space inversion transformation ${\mathcal
P}$, ${\mathbf e}$ changes sign as a vector and ${\mathbf b}$
remains invariant as a pseudovector. Let us now consider time
inversion, that is better characterized as inversion in the
direction of movement because we don't have the ``time'' variable
on our photon tensor. The direction of movement, ${\mathbf k}$,
is given by ${\mathbf e}\times{\mathbf b}$; therefore under the
time inversion transformation ${\mathcal T}$, either ${\mathbf
e}$ or ${\mathbf b}$ must change sign. Since ${\mathcal T}$ must
be different from ${\mathcal P}$ we choose that ${\mathbf b}$
changes sign whereas ${\mathbf e}$ remains invariant. In order to
fix the transformation properties of ${\mathbf e}$ and ${\mathbf
b}$ under the charge conjugation transformation ${\mathcal C}$,
we require that ${\mathcal P}{\mathcal T}{\mathcal C}$ should be
the identity; therefore both ${\mathbf e}$ and ${\mathbf b}$ must
change sign under charge conjugation. We have then the
transformation properties, including the duality transformation
${\mathcal D}$, given by
\begin{equation}\label{tranf}
  {\mathcal P}:\left\{\begin{array}{cl}
    {\mathbf e} &\rightarrow -{\mathbf e} \\
    {\mathbf b} &\rightarrow {\mathbf b} \\
  \end{array}\right .
\ ,\
 {\mathcal T}:\left\{\begin{array}{cl}
    {\mathbf e} &\rightarrow {\mathbf e} \\
    {\mathbf b} &\rightarrow -{\mathbf b} \\
  \end{array}\right .
\ ,\
 {\mathcal C}:\left\{\begin{array}{cl}
    {\mathbf e} &\rightarrow -{\mathbf e} \\
    {\mathbf b} &\rightarrow -{\mathbf b} \\
  \end{array}\right .
\ ,\
 {\mathcal D}:\left\{\begin{array}{cl}
    {\mathbf e} &\rightarrow -{\mathbf b} \\
    {\mathbf b} &\rightarrow {\mathbf e} \\
  \end{array}\right .
\ .
\end{equation}

Let us summarize all that we have learnt about the classical
photon: \emph{in any reference frame, we can visualize a positive
or negative helicity photon of energy $E$ and spin $\hbar$
propagating with speed $c$ in a direction given by a unit vector
${\mathbf k}$ as a unit vector ${\mathbf \hat{e}}$ rotating
clockwise or counterclockwise in a plane orthogonal to ${\mathbf
k}$ with frequency $\omega=E/\hbar$. In the same plane we have
another unit vector ${\mathbf \hat{b}}={\mathbf k}\times{\mathbf
\hat{e}}$ and with the vectors ${\mathbf e}=\omega{\mathbf
\hat{e}}$ and ${\mathbf b}=\omega{\mathbf \hat{b}}$ we can build
the photon tensor $f^{\mu\nu}$ whose Lorentz transformations
provide the description of the photon in other reference frames.}
\section{QUANTUM MECHANICAL TREATMENT OF ONE PHOTON}
For the quantum mechanical treatment of the photon we must first
define a Hilbert space whose elements are possible states of one
photon and where the photon observables act as hermitian
operators. There are clear empirical facts, for instance, the
emission of one photon in atomic transitions $p\!\rightarrow\!\!
s$, that indicate that the photon has spin one. Therefore the
description of spin states of the photon requires a three
dimensional Hilbert space ${\mathcal H}^{S}$. On the other side,
the kinematic description of the photon, that is, its movement in
physical space, requires an infinite dimensional Hilbert space
${\mathcal H}^{K}$. The states of a photon are then elements of
the Hilbert space
\begin{equation}\label{hil}
  {\mathcal H}={\mathcal H}^{S}\otimes{\mathcal H}^{K}\ .
\end{equation}
Photon observables are represented in this space by hermitian
operators of the form $\textbf{1}\otimes{\mathbf R}$ for
position, $\textbf{1}\otimes{\mathbf P}$ for momentum or
${\mathbf S}\otimes\textbf{1}$ for spin. Having clarified this,
we will use the simpler notation ${\mathbf R}$, ${\mathbf P}$ and
${\mathbf S}$ to denote the operators.

For all massive particles, the kinematic and the spin degrees of
freedom are decoupled and any combination of spin states with
kinematic states are acceptable. For the photon, however, we have
seen in Eq.(\ref{constr}) that spin must be collinear with
momentum, therefore states corresponding to null projection of
spin in the direction of momentum must be excluded. This means
that spin states are limited to a two dimensional subspace of
${\mathcal H}^{S}$ spanned by two states $\{\chi_{+},\chi_{-}\}$
corresponding to helicities $\pm 1$. That is,
\begin{equation}\label{heli}
    ({\mathbf k}\cdot{\mathbf S})\ \chi_{\pm}=\pm\hbar\chi_{\pm}\ ,
\end{equation}
where ${\mathbf k}$ is a unit vector (not an operator) in the
direction of ${\mathbf P}$. This two dimensional subspace is not
fixed in ${\mathcal H}^{S}$ because it is coupled to the
direction of momentum in physical space and this requires a
special treatment of spin adequate to this situation. It is usual
to choose the $z$ axis as the quantization axis for angular
momentum making the operator $S_{z}$ diagonal in such a
representation. For the case of a photon where spin is collinear
with momentum such a choice is not convenient. We will therefore
adopt a representation for the spin operators more adequate for
the implementation of the constraint ${\mathbf S}\times {\mathbf
P} =0$ with arbitrary direction of the momentum. In this
representation, the three spin operators $S_{k}$, $k=1,2,3$,
satisfying the commutation relations
$[S_{j},S_{k}]=i\hbar\varepsilon_{jkl}S_{l}$, are given by the
matrices
\begin{equation}\label{spinmat}
    S_{x}=\hbar\left(%
\begin{array}{rrr}
  0 & 0 & 0 \\
  0 & 0 & -i \\
  0 & i& 0 \\
\end{array}%
\right)\ ,\
  S_{y}=\hbar\left(%
\begin{array}{rrr}
  0 & 0 & i \\
  0 & 0 & 0 \\
  -i & 0& 0 \\
\end{array}%
\right)\ ,\
  S_{z}=\hbar\left(%
\begin{array}{rrr}
  0 & -i & 0 \\
  i & 0 & 0 \\
  0 & 0 & 0 \\
\end{array}%
\right)\ ;
\end{equation}
that is, with the matrix elements given by
\begin{equation}\label{spinmat1}
  ( S_{j} )_{kl}=-i\hbar\varepsilon_{jkl}\ .
\end{equation}
Applying the useful identity
$\varepsilon_{jkl}\varepsilon_{jrs}=\delta_{kr}\delta_{ls}-
\delta_{ks}\delta_{lr}$ one can check that the commutation
relations are satisfied. Now we can write Eq.(\ref{heli}) in
matrix form and find the eigenvectors
\begin{equation}\label{eigev}
   \chi_{\pm}=\frac{1}{2\sqrt{1-k_{x}k_{y}-k_{y}k_{z}-k_{z}k_{x}}}\left(%
\begin{array}{c}
  1- k_{x}(k_{x}+k_{y}+k_{z})\pm i(k_{y}-k_{z})\\
1- k_{y}(k_{x}+k_{y}+k_{z})\pm i(k_{z}-k_{x})\\
1- k_{z}(k_{x}+k_{y}+k_{z})\pm i(k_{x}-k_{y}) \\
\end{array}%
\right)\ ,
\chi_{0}=\left(%
\begin{array}{c}
 k_{x}\\
 k_{y}\\
 k_{z} \\
\end{array}%
\right)\ ,
 \end{equation}
where we have included the eigenvector $\chi_{0}$ incompatible
with the transversality constraint given in Eq.(\ref{constr}).

The two dimensional subspace of ${\mathcal H}^{S}$ containing the
spin states is orthogonal (and therefore the name
\emph{transversality} given to the constraint) to the Hilbert
space element $\chi_{0}$ that has, \emph{numerically}, the same
components as the vector ${\mathbf k}$ in physical space. It is
important not to confuse these two objects: $\chi_{0}$ belongs to
the Hilbert space of states whereas ${\mathbf k}$ belongs to
physical space. This confusion of the three dimensional Hilbert
space of states with the three dimensional physical space is the
source of an erroneous identification of Maxwell's equations with
Schr{\"o}dinger's equation. Let us look in more details at this
interesting but, alas, erroneous argument.
\section{ON SCHR{\"O}DINGER AND MAXWELL EQUATIONS}
Using the mathematical tools developed for quantum mechanics, it
has been shown that Maxwell's equations can be written as an
evolution equation for a spinor\cite{MxSp,birul}. However it
should be clear that this spinorial equation remains a
\emph{classical} equation for the \emph{classical}
electromagnetic field disguised as a quantum mechanical equation
and we should resist the temptation of an interpretation of this
equation as some ``Schr{\"o}dinger's equation'' for some quantum
system like the photon. In this section we will see that the
attempted derivation of Maxwell's equations from the photon's
Schr{\"o}dinger equation is erroneous.

Let us consider a photon state $\{\Psi_{j}\}\in{\mathcal
H}^{S}\otimes{\mathcal H}^{K}$ where the index $j=1,2,3$ denotes
the components of the state in the three dimensional space
${\mathcal H}^{S}$. Its time evolution will be determined by a
hamiltonian operator such that it leaves invariant a photon state
$\chi_{\pm}^{{\mathbf k}}\otimes\phi_{{\mathbf p}}$, corresponding to
the momentum eigenvalue ${\mathbf p}$ in the direction ${\mathbf
k}={\mathbf p}/|{\mathbf p}|$, with helicity $\pm 1$ and energy
$E=c|{\mathbf p}|$. One candidate to achieve this, is the
operator
\begin{equation}\label{schr0}
 {\mathbf S}\cdot{\mathbf P}={\mathbf k}\cdot{\mathbf S}\otimes|{\mathbf P}|
\ ,
\end{equation}
where the operator $|{\mathbf P}|$ is such that $|{\mathbf
P}|\phi_{{\mathbf p}}=|{\mathbf p}|\phi_{{\mathbf p}}$. Indeed we
have
\begin{equation}\label{schr01}
 {\mathbf S}\cdot{\mathbf P}\ \chi_{\pm}^{{\mathbf
k}}\otimes\phi_{{\mathbf p}}=\pm \frac{\hbar E}{c}\
\chi_{\pm}^{{\mathbf k}}\otimes\phi_{{\mathbf p}}\ .
\end{equation}
The hamiltonian chosen is then
\begin{equation}\label{schr1}
 H=\frac{c}{\hbar}{\mathbf S}\cdot{\mathbf P}\ ,
\end{equation}
however this operator is not positive, as can be seen in
Eq.\ref{schr01}, and therefore the minus sign should be correctly
interpreted as a photon with positive energy but negative
helicity. With this, the time evolution of a general photon state
will be given by
\begin{equation}\label{schr3}
   i\hbar\partial_{t}\Psi_{j}=(H)_{jk}\Psi_{k}\ ,
\end{equation}
where we have explicitly written the hamiltonian as a $3\times 3$
matrix in ${\mathcal H}^{S}$ (whose components are operators in
${\mathcal H}^{K}$). That is,
\begin{equation}\label{schr4}
  i\hbar\partial_{t}\Psi_{j}=\frac{c}{\hbar}
({\mathbf S}\cdot{\mathbf P})_{jk}\Psi_{k}\ .
\end{equation}
We can now specialize this equation for a particular
representation of the Hilbert space. For the spin part we choose
the representation where spin is given by Eq.(\ref{spinmat1}) and
for the kinematic part ${\mathcal H}^{K}$ we choose the space of
square integrable functions of ${\mathbf r}$, ${\mathcal L}_{2}$
where the momentum operator is given by the derivative operator
${\mathbf P}=-i\hbar\nabla$. In this way we obtain what we may
call \emph{Schr{\"o}dinger's equation for the photon}
\begin{equation}\label{schr5}
\frac{i}{c}\partial_{t}\Psi_{j}(t,{\mathbf
r})=\varepsilon_{jlk}\partial_{l} \Psi_{k}(t,{\mathbf r})\ .
\end{equation}
Notice that in this equation the Planck constant $\hbar$ does not
appear although it is essentially a quantum mechanical equation.
This can be interpreted as an indication that this equation does
not have a classical limit $\hbar\rightarrow 0$. This equation is
correct but we will now see how, doing some illegal moves, we can
get from it Maxwell's equations. This intended derivation
contains an important conceptual and mathematical error
consisting in identifying the three dimensional Hilbert space of
states with the three dimensional physical space and to equate
$\Psi_{j}$ with the components of the electromagnetic vector
fields ${\mathbf E}$ and ${\mathbf B}$ according to
\begin{equation}\label{schr6}
  \Psi_{j}(t,{\mathbf r})=E_{j}(t,{\mathbf r})+iB_{j}(t,{\mathbf
r})\ \ \ (?)\ .
\end{equation}
Replacing above and separating the real and imaginary part, we
obtain the first two of Maxwell's equations
\begin{eqnarray}\label{mx1}
  -\varepsilon_{jlk} \partial_{l}E_{k}&=& \frac{1}{c}\partial_{t}B_{j}\ , \\
  \varepsilon_{jlk} \partial_{l}B_{k}&=& \frac{1}{c}\partial_{t}E_{j}\ , \\
 \partial_{k}E_{k}&=&0\ ,\\
\partial_{k}B_{k}&=&0 \ .
\end{eqnarray}
The last two equations are obtained repeating the same error: the
photon state must be orthogonal to the state $\chi_{0}$ of
${\mathcal H}^{S}$ given in Eq.(\ref{eigev}), that is,
$k_{k}\Psi_{k}=0$. Now replacing the \emph{real numbers} $k_{k}$
by the \emph{operators} $P_{k}$ in the ${\mathcal L}_{2}$
representation, another error,  we get $\partial_{k}\Psi_{k}=0$
that with the misidentification of Eq.(\ref{schr6}) leads to the
last two Maxwell's equations.

Certainly, the manipulations shown can not be considered to be a
derivation of Maxwell's equations from the photon Schr{\"o}dinger's
equation. The intended identification in Eq.(\ref{schr6}) of the
state function of a quantum system with a field having physical
existence, that is carrying energy an momentum in physical space,
was considered in the early days of quantum mechanics by
Schr{\"o}dinger himself; however he had to abandon such an
interpretation mainly because of the failure of the
interpretation for a system of many particles. Indeed, the state
function for $N$ particles depends on $3N$ coordinates and
therefore it can not be considered to be a field in three
dimensional physical space. It is today well known that the
quantum mechanical state can not be interpreted as an objectively
existing field carrying energy and momentum as the
electromagnetic field does. Indeed, if the quantum state were a
field carrying energy and momentum, then well established quantum
effects like nonlocality, entanglement, teleportation and others
would imply unacceptable violations of relativity.

Another useful representation of Eq.(\ref{schr4}) results from
the observation that the equation will take its simplest form in
the Hilbert space where the momentum operator has the simplest
expression and this is, of course, the space of square integrable
functions of ${\mathbf p}$ where the momentum operator amounts
simply to a multiplication by ${\mathbf p}$. Taking, as before,
for ${\mathcal H}^{S}$ the representation where spin is given by
Eq.(\ref{spinmat1}) and ${\mathcal L}_{2}$ for the kinematic part
${\mathcal H}^{K}$, we obtain the photon Schr{\"o}dinger's equation
in momentum representation
\begin{equation}\label{schr7}
\frac{\hbar}{c}\partial_{t}\Psi_{j}(t,{\mathbf
p})=-\varepsilon_{jlk}p_{l} \Psi_{k}(t,{\mathbf p})\ .
\end{equation}
In spite of the formal similarity, the right hand side of this
equation \emph{should not} be written as a ``vector product''
${\mathbf p}\times{\mathbf \Psi}$ because the two ``vectors''
belong to different spaces: ${\mathbf p}$, the momentum
eigenvalue, is in physical three dimensional space and ${\mathbf
\Psi}$ is in the three dimensional Hilbert space ${\mathcal
H}^{S}$. In Eqs.(\ref{schr7}) and (\ref{schr5}) we have used the
same letter $\Psi$ to denote different functions (actually
related by Fourier transformation) but this should cause no
confusion.

We can now investigate the stationary state solutions to
Eq.(\ref{schr4}), that is, solutions corresponding to a fixed
value of the energy $E$ of the form
\begin{equation}\label{stat}
 \Psi_{j}=\exp(-\frac{i}{\hbar}Et)\Phi_{j,E }
\end{equation}
where $\Phi_{j,E }$ is the solution of the time independent
equation
\begin{equation}\label{schr8}
({\mathbf S}\cdot{\mathbf P})_{jk}\Phi_{k,E}=\frac{E}{c}\hbar\
\Phi_{j,E }\ .
\end{equation}
Again, this equation will take its simplest form in the momentum
representation. With respect to the spin part, the equation above
is essentially the same as Eq.(\ref{heli}) and therefore the spin
part of the solution is given by the helicity states given in
Eq.(\ref{eigev}). The kinematic part of the solution must relate
the momentum eigenvalue ${\mathbf p}$ with the energy $E$.
Therefore we have
\begin{equation}\label{schr9}
\Phi_{j,E}({\mathbf p})=\ \chi_{\pm}^{{\mathbf
k}}\otimes\delta\left(|{\mathbf p}|-\frac{E}{c}\right)\ ,
\end{equation}
Where ${\mathbf k}={\mathbf p}/|{\mathbf p}|$. As said before,
the momentum representation is the most adequate one in order to
describe the quantum state of the photon because in this case the
transversality constraint, that is, the condition that the spin
and momentum must be collinear, is most evident. In order to
obtain the position representation of the photon energy
eigenstates we must take the Fourier transform of the equation
above that does not have a simple expression. For this reason, in
the position representation it is more convenient to describe the
photon in momentum eigenstate rather than in energy eigenstates.
These states are $\chi_{\pm}^{{\mathbf k_{0}}}\otimes\phi_{{\mathbf
p_{0}}}$ with ${\mathbf k_{0}}={\mathbf p_{0}}/|{\mathbf p_{0}}|$
corresponding to a photon with definite momentum ${\mathbf
p_{0}}$ and helicity $\pm 1$. These states are also eigenstates
of the hamiltonian with degenerate energy eigenvalue
$E=c|{\mathbf p_{0}}|$. The kinematic part $\phi_{{\mathbf
p_{0}}}$ in the momentum representation is given by
\begin{equation}\label{momrep}
\phi_{{\mathbf p_{0}}}({\mathbf p})=\delta({\mathbf p}-{\mathbf
p_{0}}) \ ,
\end{equation}
and in position representation they are
\begin{equation}\label{posrep}
\phi_{{\mathbf p_{0}}}({\mathbf
r})=\frac{1}{(\sqrt{2\pi}\hbar)^{3}}\exp(\frac{i}{\hbar}{\mathbf
p_{0}}\cdot{\mathbf r}) \ ,
\end{equation}
showing that \emph{one} photon behaves as a wave in these states.
Notice that this is the first appearance of a wave related to the
photon. We will later see the relation of this wave with the
macroscopic electromagnetic waves.
\section{COMPLEMENTARITY FOR THE PHOTON}
The idea of complementarity, introduced by Bohr in quantum
mechanics, is a generalization of the observed wave-particle
duality in the behaviour of matter. According to complementarity,
every description or observation of a physical system involves a
set of observables that necessarily excludes other observables.
The simultaneous and exact treatment of all observables of the
system is not possible. As as metaphor for complementarity one
can think that the observation of a human face must be
necessarily done from only one particular point of view. Every
perspective excludes other possible perspectives. A
philosophically realist physicist postulates the objective
existence of physical reality that may be in different quantum
states corresponding to all possible complementary descriptions
or observations, in the same way that the tree dimensional real
existence of a human face generates all possible two dimensional
perspectives. The marvellous invention of Picasso, that combines
mutually exclusive perspectives of a human face in one single
image is, in this sense, a representation of reality more
faithful than a conventional photograph showing only one
perspective.

As it happens with any particle, the photon also exhibits the
wave-particle duality in different experimental situations. It
behaves like a particle in the photoelectric effect and in
Compton scattering and it behaves like a wave in diffraction
experiments. The photon exhibits particle like behaviour when it
is in a localized quantum state $\varphi_{{\mathbf r}}$,
eigenvector of the position operator ${\mathbf R}$ and it will
behave as a wave when it is in state $\phi_{{\mathbf p}}$
eigenstate of the momentum operator ${\mathbf P}$. Both dual
states, or any other, are equally valid; however the
transversality constraint requires a well defined momentum and
therefore it introduces a preference for momentum eigenstates. In
order to impose the transversality constraint in an arbitrary
state we must expand it on the basis $\{\phi_{{\mathbf p}}\}$ of
${\mathcal H}^{K}$ and every component of it will determine the
two dimensional subspace of ${\mathcal H}^{S}$ that contains the
spin state. An interesting remark is that in such a state, spin
and momentum are entangled because every momentum eigenvalue is
associated with a spin superposition. An arbitrary state for one
photon is then
\begin{equation}\label{gensta}
   \Psi=\int\!\!\! d^{3}{\mathbf p}\ C({\mathbf p}) \left(
\alpha({\mathbf p})\chi_{+}^{{\mathbf p}}+\beta({\mathbf
p})\chi_{-}^{{\mathbf p}} \right) \otimes\phi_{{\mathbf p}}\ .
\end{equation}
where $C({\mathbf p})$, $\alpha({\mathbf p})$ and $\beta({\mathbf
p})$ are the coefficients that determine the state, and
$\chi_{\pm}^{{\mathbf p}}$ are the helicity states, given by
Eq.(\ref{eigev}) with ${\mathbf k}={\mathbf p}/|{\mathbf p}|$.

It is necessary to emphasize that the wave-particle duality for
one photon, as for any other single material particle,
corresponds to different quantum states associated to
noncommuting observables of position and momentum, because there
is some confusion proposing that the photon is the dual,
particle-like, partner of the wave-like electromagnetic field.
The electromagnetic field is not a quantum state for a particle
exhibiting wave like behaviour. The electromagnetic field is a
macroscopic observable of a system composed by a large number, or
more precisely, an undetermined number of photons (we will see
however one special case where all these photons are
Bose-Einstein condensed in a single photon state). The explicit
construction of the electromagnetic fields in terms of an
ensemble of photons will be treated in another work.\cite{foton3}
\section{CONCLUSIONS}
In this work devoted to the attempt to understand the quantum of
electromagnetic radiation, we have described the single free
photon, first as a classical relativistic particle that requires
for its description an antisymmetric tensor, the photon tensor.
In this description we developed an image of the photon as an
essentially two dimensional object with its elements of physical
reality on a plane orthogonal to the direction of propagation. As
a consequence of special relativity, not of quantum mechanics,
the energy of the photon is related to a frequency that we
associate to a clockwise or counterclockwise rotation in the
plane of the photon. This rotation is simultaneously related to
the spin of the photon. It is shown that this image of the photon
is compatible with the empirical fact that the energy depends
linearly on the frequency of rotation and the spin is constant
independent of the frequency.

The quantum mechanical description of a free photon was given in
similar lines as the description of any particle, but with the
complication that spin and momentum are coupled, or entangled by
the transversality constraint arising from the massless character
of the photon. This condition implies a preference for momentum
eigenstates compared to localized states. It was shown that the
quantum states of a photon can not be associated to the
electromagnetic fields as could be suggested by a mistaken
derivation of Maxwell's equations from the photon Schr{\"o}dinger's
equation.

Finally, in these attempts to understand the quantum of
electromagnetic radiation, the basic assumption was made
considering that the photons are the primary ontology, the
building blocks, of electromagnetism and that the electromagnetic
fields are a macroscopic construct emergent as a collective
effect of an ensemble of photons. The details of the construction
of the electromagnetic fields with  a collection of photons will
be the subject of another work.\cite{foton3}
\section{APPENDIX}
In this section we build a toy model for a photon. For this, we
show that it is possible to have a rotating  energy distribution
in a plane, such that the total energy $E$ is proportional to the
frequency of rotation $\omega_{0}$ and the associated angular
momentum $S$ is constant, i.e. independent of $\omega_{0}$.
Furthermore $E= \omega_{0}S$.

Let $\varepsilon(r)$ be the energy distribution in a plane, where
$r$ is the radial coordinate. The total energy is then
\begin{equation}\label{totener}
  E=\int_{0}^{\infty}\!\!\!\!\! dr\: 2\pi r \:\varepsilon(r) \ .
\end{equation}
The energy distribution is rotating with an angular frequency
$\omega_{0}$ at the center $r=0$. The rotation is not rigid and
at any distance $r$ we have a local angular velocity $\omega(r)$
and a tangential velocity $v(r)$ related by $v(r)=r\omega(r)$. We
must determine the local tangential and angular velocity in a way
consistent with relativity, that is, for increasing $r$ the
velocity should not exceed Einstein's constant $c$. For this, let
us consider two points at $r$ and $r+dr$. The velocity $v(r+dr)$
is given by the relativistic combination of the velocity $v(r)$
and the radial increase in velocity $dr\:\omega(r)$. That is,
\begin{equation}\label{rot}
    v(r+dr)=\frac{v(r)+dr\:\omega(r)}{1+v(r)\:dr\:\omega(r)/c^{2}} \ .
\end{equation}
That is, in terms of angular velocity alone,
\begin{equation}\label{rot1}
    \omega(r+dr)=\frac{\omega(r)}{1+r\:\omega^{2}(r)\:dr/c^{2}} \ .
\end{equation}
With this we build the derivative
\begin{equation}\label{rot2}
    \frac{d\omega(r)}{dr}=-\frac{\omega^{3}(r)}{c^{2}}\:r\ ,
\end{equation}
and integrating we get
\begin{equation}\label{rot3}
   -c^{2} \int^{\omega(r)}_{\omega_{0}}\!\!\!\!\!\!\!d\omega\:
   \omega^{-3}=\int^{r}_{0}\!\!\!\!\!dr\:r \ .
\end{equation}
That is,
\begin{equation}\label{rot4}
   \omega(r)=\frac{\omega_{0}}{\sqrt{(\frac{\omega_{0}r}{c})^{2}+1}}\ ,
\end{equation}
and the velocity is
\begin{equation}\label{rot5}
 v(r)=r\:\omega(r)=\frac{\omega_{0}r}{\sqrt{(\frac{\omega_{0}r}{c})^{2}+1}}\
 .
\end{equation}
With increasing $r$, the angular velocity decreases and the
tangential velocity increases approaching the constant $c$.

In order to determine the angular momentum associated to the
rotating energy density we consider that an element of area at
position $r$ will have an \emph{effective} mass
$\varepsilon(r)/c^{2}$ and a tangential velocity $v(r)$ and
therefore we have momentum density
$\pi(r)=\frac{\varepsilon(r)}{c^{2}}\gamma(r)\:v(r)$ where
$\gamma(r)$ is the Lorentz factor corresponding to the velocity
$v(r)$. After some simple algebra we get
\begin{equation}\label{Angmom}
    \pi(r)=\frac{\varepsilon(r)}{c^{2}}\omega_{0}\:r\ .
\end{equation}
Now, multiplying the momentum density by the radial distance and
integrating we get the total angular momentum
\begin{equation}\label{Angmom1}
    S=\int_{0}^{\infty}\!\!\!\!\! dr\: 2\pi r^{3} \:
    \frac{\varepsilon(r)}{c^{2}}\omega_{0}\ .
\end{equation}

Now, considering Eqs.(\ref{totener},\ref{Angmom1}) we can find a
frequency depending energy distribution $\varepsilon(r)$ such
that $S=\hbar$ and $E=\omega_{0}\hbar$. There are many energy
distributions that fulfill these conditions providing several toy
models for the photon. The simplest example is a rotating disk
of radius $R=\frac{\lambda}{\sqrt{2}\pi}$ with constant energy
distribution
\begin{equation}\label{disk}
    \varepsilon(r)=
\left\{%
\begin{array}{ll}
    \frac{\hbar \omega^{3}_{0}}{2\pi c^{2}}, & \hbox{for } r\leq\frac{\lambda}{\sqrt{2}\pi}\\
    0, & \hbox{for }r>\frac{\lambda}{\sqrt{2}\pi} \\
\end{array}%
\right. \ .
\end{equation}
In another example the energy distribution is not constant and
has support in an annular region with internal radius $R_{1}$ and
external radius $R_{2}$:
\begin{equation}\label{ring}
    \varepsilon(r)=\left\{%
\begin{array}{ll}
    0, & \hbox{for }r<R_{1} \\
    \frac{K}{r^{3}}, & \hbox{for } R_{1}\leq r\leq R_{2} \\
    0, & \hbox{for } r>R_{2}\\
\end{array}%
\right.
\end{equation}
where the $K,R_{1},R_{2}$ are given in terms of a parameter $k$
by
\begin{equation}\label{ring1}
  R_{1}=\frac{\lambda}{2\pi}\left(\sqrt{k^{2}+1}-k\right)\ ,\
 R_{2}-R_{1}=\frac{\lambda}{\pi}k \ ,\
 K=\frac{\hbar c}{4\pi k} \ .
\end{equation}
When $k\rightarrow 0$ the annular region becomes a ring of radius
$\lambda/(2\pi)$. In these two examples the total surface where
the energy doesn't vanish, that is, the photon total cross
section, depends on the photon energy as $\sigma_{T}\propto
E^{-2}$.

Another last example for a toy photon is provided by an energy
distribution along a rotating segment or string. In this case the
expressions for energy and spin are
\begin{equation}\label{totenerlin}
  E=\int_{-\infty}^{\infty}\!\!\!\!\! dx\:\varepsilon(x) \ ,
\end{equation}
\begin{equation}\label{Angmom1lin}
    S=\int_{-\infty}^{\infty}\!\!\!\!\! dx\:x^{2} \:
    \frac{\varepsilon(x)}{c^{2}}\omega_{0}\ ,
\end{equation}
and an energy distribution
\begin{equation}\label{string}
    \varepsilon(x)=
\left\{%
\begin{array}{ll}
    \frac{\hbar \omega^{2}_{0}}{2\sqrt{3}c}, & \hbox{for }
    |x|\leq\frac{\sqrt{3}}{2\pi}\lambda\\
    0, & \hbox{otherwise } \\
\end{array}%
\right. \ ,
\end{equation}
will have $E=\hbar \omega_{0}$ and $S=\hbar$.
\begin{acknowledgements}
I would like to thank H. M{\'a}rtin, O. Sampayo and A. Jacobo for
challenging discussions. This work received partial support from
``Consejo Nacional de Investigaciones Cient\'{\i}ficas y
T\'ecnicas'' (CONICET), Argentina.
\end{acknowledgements}

\end{document}